%%%%%%%%%%%%%%%%%%%%%%%%%%%%%%%%%%%%%%%%%%%%%%%%%%%%%%%%%%%%
%%%%%%%%%  Strings in plane-fronted gravitational waves   %%%%%%%%%%
%%%%%%%%%%%%%% Corrected the 19-Nov-93 %%%%%%%%%%%%%%%%%%%%%%
%%%%%%%%%%%%%%%%%%%%%%%%%%%%%%%%%%%%%%%%%%%%%%%%%%%%%%%%%

%\magnification1200
\font\BBig=cmr10 scaled\magstep2
\font\BBBig=cmr10 scaled\magstep3

%%%%%%%%%%%%%%%%%%%%%%%%%%%%%%%%%%%%%%%%%%%
%%%%%%%%%%%%%% the title %%%%%%%%%%%%%%%%%%
%%%%%%%%%%%%%%%%%%%%%%%%%%%%%%%%%%%%%%%%%%%

\def\title{
{\bf\BBBig
\centerline{Strings in plane-fronted gravitational waves}
}}%%%%% for the front page

\def\foot#1{
\footnote{($^{\the\foo}$)}{#1}\advance\foo by 1
} %%%%% foot(notes
\def\ccr{\cr\noalign{\medskip}}

%%%%%%%%%%%%%%%%%%%%%%%%%%%%%%%%%%%%%%%%%%%
%%%%%%%%%%%%% the author(s) %%%%%%%%%%%%%%%
%%%%%%%%%%%%%%%%%%%%%%%%%%%%%%%%%%%%%%%%%%%

\def\authors{
\centerline{C.~DUVAL\foot{
Centre de Physique Th\'eorique, CNRS 
Luminy, Case 907, 
F-13288 MARSEILLE Cedex 9 (France).
UMR 6207 du CNRS associ\'ee aux 
Universit\'es d'Aix-Marseille I et II
et Universit\'e du Sud Toulon-Var. Laboratoire 
affili\'e \`a la FRUMAM-FR2291.
e-mail: duval@cpt.univ-mrs.fr},
Z.~HORV\'ATH\foot{Institute for Theoretical Physics,
E\"otv\"os University, H--1088 BUDAPEST,
Puskin u.~5-7 (Hungary);
e-mail: zalanh@ludens.elte.hu}
and
P.~A.~HORV\'ATHY\foot{Laboratoire de Math\'ematiques
et de Physique Th\'eorique,
Universit\'e de Tours, Parc de Grandmont,
F--37200 TOURS (France);
e-mail: horvathy@univ-tours.fr}}
}

\def\runningauthors{
Duval, Horv\'ath, 
Horv\'athy
} %%%%% for the header

\def\runningtitle{
Strings in gravitational waves}

%%%%%%%%%%%%%%%%%%%%%%%%%%%%%%%%%%%%%%%%%%%
%%%%%%%%%%%%%% the metrics %%%%%%%%%%%%%%%%
%%%%%%%%%%%%%%%%%%%%%%%%%%%%%%%%%%%%%%%%%%%

%\hsize = 12.5cm %%%%% text width
%\hoffset = 1cm %%%%% hor printing
\voffset = 1cm %%%%% ver printing
\baselineskip = 14pt %%%%% line spacing

\headline ={
\ifnum\pageno=1\hfill
\else\ifodd\pageno\hfil\tenit\runningtitle\hfil\tenrm\folio
\else\tenrm\folio\hfil\tenit\runningauthors\hfil
\fi\fi
} %%%%% header

\nopagenumbers
\footline = {\hfil} %%%%% footer

%%%%%%%%%%%%%%%%%%%%%%%%%%%%%%%%%%%%%%%%%%%
%%%%%%%%%%%%% greek boldface %%%%%%%%%%%%%%
%%%%%%%%%%%%%%%%%%%%%%%%%%%%%%%%%%%%%%%%%%%

\font\tenb=cmmib10
\newfam\bsfam
 
\textfont\bsfam=\tenb
\mathchardef\betab="080C
\mathchardef\xib="0818
\mathchardef\omegab="0821
\mathchardef\deltab="080E
\mathchardef\epsilonb="080F
\mathchardef\pib="0819
\mathchardef\sigmab="081B
\mathchardef\bfalpha="080B
\mathchardef\bfbeta="080C
\mathchardef\bfgamma="080D
\mathchardef\bfomega="0821
\mathchardef\zetab="0810

%%%%%%%%%%%%%%%%%%%%%%%%%%%%%%%%%%%%%%%%%%%%%%%%%%%%%%
%%%%%%%%%%%%%%%%%% some definitions %%%%%%%%%%%%%%%%%%
%%%%%%%%%%%%%%%%%%%%%%%%%%%%%%%%%%%%%%%%%%%%%%%%%%%%%%

\def\and{\qquad\hbox{and}\qquad}
\def\where{\qquad \hbox{where}\qquad}
\def\kikezd{\parag\underbar}

\def\IR{{\bf R}}

\def\math#1{\mathop{\rm #1}\nolimits}
\def\const{\math{const}}

\def\smallcirc{{\raise 0.5pt \hbox{$\scriptstyle\circ$}}}
\def\smallover#1/#2{\hbox{$\textstyle{#1\over#2}$}}
\def\2{{\smallover 1/2}}

\def\ccr{\cr\noalign{\medskip}}
\def\parag{\hfil\break} %%%%% paragraph
\def\={\!=\!}
\def\-{\!-\!}
\def\tilde{\widetilde}
\def\hat{\widehat}
%%%%%%%%%%%%%%%%%%%%%%%%%%%%%%%%%%%%%%%%%%%
%%%%%%%%%%%%%%% numberings %%%%%%%%%%%%%%%%
%%%%%%%%%%%%%%%%%%%%%%%%%%%%%%%%%%%%%%%%%%%

\newcount\ch %%%%% ch(apters
\newcount\eq %%%%% eq(uations
\newcount\foo %%%%% foo(tnotes
\newcount\ref %%%%% ref(erences

\def\chapter#1{
\parag\eq = 1\advance\ch by 1{\bf\the\ch.\enskip#1}
}

\def\equation{
\leqno(\the\ch.\the\eq)\global\advance\eq by 1
}

\def\reference{
\parag [\number\ref]\ \advance\ref by 1
}

\ch = 0 %%%%% global init ch(apter
%%%%% eq is set to 1 by \chapter
\foo = 1 %%%%% global init foo(tnote
\ref = 1 %%%%% global init ref(erence
%%%%%%%%%

%%%%%%%%%%%%%%%%%%%%%%%%%%%%%%%%%%%%%%%%%%%
%%%%%%%%%%%%%%%% the text %%%%%%%%%%%%%%%%%
%%%%%%%%%%%%%%%%%%%%%%%%%%%%%%%%%%%%%%%%%%%

\title
\vskip .5cm
\authors
\vskip .12in

\parag
{\bf Abstract.} 
Brinkmann's plane-fronted gravitational waves with parallel rays
---~shortly pp-waves~--- are shown to provide, under suitable
conditions, exact string vacua at all orders of the sigma-model
perturbation expansion.

\vskip .3cm
\noindent{September 1993}. 
Mod. Phys. Lett. {\bf A8}: 3749-3756, 1993.

\chapter{Introduction}

The condition for the vanishing of the conformal
anomaly for a bosonic string in a curved space is
expressed, in $\sigma$-model perturbation theory,
as $R_{\mu\nu}+\ldots=0$, where the ellipsis denotes
rank two tensors constructed from
derivatives and powers of the Riemann tensor~[1,2].
An example for which all
such correction terms vanish is provided by
`ordinary' plane-fronted gravitational waves,
$$
d{\bf x}^2-2dudv+K(u,{\bf x})du^2,
\equation
$$
where ${\bf x}$ belongs to the $(D-2)$-dimensional flat
transverse space and $u$ and $v$ are two additional
light-cone coordinates.
Then all higher-order correction terms vanish due to
the special form of the Riemann tensor, and the vanishing
anomaly condition reduces to 
the vacuum Einstein equation $R_{\mu\nu}\=0$ i.e. to 
$$
\Delta K=0,
\equation
$$
where $\Delta\equiv\Delta_{D-2}$ is the (flat) transverse
Laplacian~[3,4].
For    
exact plane waves, 
$K\=\sum_{ij}{K_{ij}(u)x^ix^j}$
where $K_{ij}$ is symmetric and traceless
(and for $D=26$), the constraint $\Delta K\=0$ enforces 
the anomaly cancellation even non-perturbatively~[3].

This result can be extended by including other massless fields,
namely a dilaton, $\Phi(u)$, and an axion, $b_{\mu\nu}$,
with only non-zero component
$b_{iu}\=\2B_{ij}(u)x^j$
where $B_{ij}$ is antisymmetric.
Higher-order terms vanish again, and the vanishing anomaly
condition is simply~[4]
$$
\Delta K+{\smallover 1/{18}}B_{ij}B^{ij}+2\ddot\Phi=0.
\equation
$$ 
In the quadratic case the anomaly again vanishes 
non-perturbatively~[5].

Another generalization was found by Rudd~[6] who has shown
the vanishing of the anomaly for a metric + dilaton system
with metric
$$
\sum_{i=1}^{D-2}[2\pi R_i(u)]^2(dx^i)^2-2dudv,
\equation
$$
provided
$$
\sum_{i=1}^{D-2}{{\ddot R_i\over R_i}}-2\ddot\Phi=0.
\equation
$$

Now both (1.1) and (1.4) are particular cases of Brinkmann's
generalized plane-fronted waves with parallel rays (shortly
pp-waves)~[7,8,9],
$$ 
{\tilde g}_{\mu\nu}dx^{\mu}dx^{\nu}
=g_{ij}(u,{\bf x})dx^idx^j-2du\Big[dv+A_i(u,{\bf x})
dx^i\Big]+k(u,{\bf x})du^2.
\equation
$$

String propagation in the metric (1.6)
has been considered by Horowitz~[10] who found
however that for a non-trivial transverse metric
the higher-order terms did not simplify.
In this Letter we show that for a {\it special choice},
Eq.~(2.3) below, the Brinkmann metric
behaves exactly as (1.1) and (1.4).

\chapter{Vanishing of the anomaly in a special Brinkmann wave.}

Let us start with the general Brinkmann metric (1.6).
Observe first that it admits a covariantly constant null vector
$(\ell^\mu)$, namely $\partial_v$.
In order to study the Weyl anomaly, let us decompose the metric
(1.6) into the sum of a  `background' metric $g_{\mu\nu}$
with the vector potential terms: setting $A_u\=k/2$ and $A_v\=0$,
(1.6) is re-written as 
$$
\tilde{g}_{\mu\nu}=g_{\mu\nu}+2\ell_{(\mu}A_{\nu)}
\where
g_{\mu\nu}dx^\mu dx^\nu=g_{ij}(u,{\bf x})dx^idx^j-2dudv.
\equation
$$
The Riemann (resp. Ricci)
tensors are found to be 
$$\left\{\eqalign{
&{\tilde R}_{\mu\nu\rho\sigma}=R_{\mu\nu\rho\sigma} 
-\ell_{[\mu}\nabla_{\nu]}F_{\rho\sigma} 
-\ell_{[\rho}\nabla_{\sigma ]}F_{\mu\nu} 
-\ell_{[\mu}F_{\nu]\,}^{\ \alpha}\ell_{[\rho}F_{\sigma]\alpha},
\ccr
&{\tilde R}_{\mu\rho}=R_{\mu\rho} 
-\ell_{(\mu}\nabla^{\nu}F_{\rho)\nu} 
-{\smallover1/4}\ell_{\mu}\ell_{\rho}F^{\nu\sigma}F_{\nu\sigma}.
\cr
}
\right.
\equation
$$
where $\nabla$ is covariant derivative with respect to the metric
$g_{\mu\nu}$, also $F_{\mu\nu}\equiv2\partial_{[\mu}A_{\nu]}$ and
$F_{\ \nu}^{\mu\ }\equiv{\tilde g}^{\mu\rho}F_{\rho\nu}$. 

A look at Eq.~(2.2) confirms that the higher order terms in the
perturbation expansion do not vanish in general~[10]. For
example,
$\tilde{R}_{\mu\rho\sigma\lambda}
{\tilde R}_{\nu}^{\ \rho\sigma\lambda}\!\neq\!0$,
etc.
The clue of Horowitz and Steif to overcome this is to express the
Riemann tensor using the covariantly constant null vector and
demand that it contain two $\ell_\mu$'s~[4,10]. We show below that
this  is satisfied by the following special choice:
$$ 
{\tilde g}_{\mu\nu}dx^{\mu}dx^{\nu}
=g_{ij}(u)dx^idx^j
-2du\Big[dv+\2e_{ij}(u)x^j dx^i\Big]+k(u,{\bf x})du^2,
\equation
$$
where $e_{ij}$ is an $u$-dependent matrix. 
Firstly, the background Riemann tensor,
$$
R_{\mu\nu\rho\sigma}
=
-2\ell_{[\nu}\ddot{g}_{\mu][\sigma}\ell_{\rho]}
+
g^{\alpha\beta}
\ell_{[\nu}\dot{g}_{\sigma]\alpha}
\ell_{[\rho}\dot{g}_{\mu]\beta},
\equation
$$
is proportional to two $\ell_\mu$'s as required. 
Secondly, the two middle terms in the Riemann tensor
in Eq.~(2.2) will contain two
$\ell$'s if no triple transverse indices arise,
$$
\nabla_iF_{jk}=0
%\qquad
\hbox{\ \ for all\ }\,i,j, k\=1,\ldots,D\-2,
\equation
$$
which is automatically satisfied by the choice (2.3).
Then the argument of Horowitz and Steif~[4] 
shows that all higher-order terms in the perturbation 
expansion vanish:

\item{(a)} In those terms with at least two
$\tilde{R}_{\mu\nu\rho\sigma}$'s
at least one of the null vectors $\ell$'s are contracted, and
vanish therefore.

\item{(b)} Terms of the form
${\tilde\nabla}^{\mu}
{\tilde\nabla}^\rho\tilde{R}_{\mu\nu\rho\sigma}$
are related to the covariant derivative of the Ricci tensor by the
Bianchi identity and hence vanish also.

\item{(c)} Terms containing 
$\tilde{R}_{\mu\nu\rho\sigma}$'s
as well as its covariant derivatives, one again has to contract 
at least one $\ell$ on an index of either the curvature tensor or 
its covariant derivative. In both cases, one gets zero.

Explicitly, the only non-vanishing components of the
background Riemann (resp. Ricci) tensors are
$$
\left\{\eqalign{
&R_{iuuj}=\2\ddot{g}_{ij}
-\smallover1/4
g^{mn}\dot{g}_{mi}\dot{g}_{nj},
\ccr
&R_{uu}=\Big(\2\ddot{g}_{ij}
-\smallover1/4
g^{mn}
\dot{g}_{mi}\dot{g}_{nj}\Big)g^{ij}.
\cr
}\right.
\equation
$$

Thus, the vacuum 
Einstein equations $\tilde{R}_{\mu\nu}\=0$ require
$$
\tilde{R}_{uu}=\Big(\2\ddot{g}_{ij}
-\smallover1/4
g^{mn}
\dot{g}_{mi}\dot{g}_{nj}\Big)g^{ij}
-\nabla^iF_{ui}
-\smallover1/4F_{ij}F^{ij}=0,
\equation
$$
and we conclude that if (2.7) holds, then we get an exact string 
solution at all orders in sigma-model perturbation theory.

\chapter{Reduction to ordinary plane wave.}

Now we explain why the special Brinkmann metric (2.3) works.
We prove in fact that (2.3) can be brought into the simple form
(1.1) by a sequence of coordinate transformations.
At each step, $x,u,v$ (resp. $X,U,V$) denote the old (resp. new)
coordinates.

\kikezd{Step 1}. The positive transverse metric
$g_{ij}\=g_{ij}(u)$ is, by assumption,
a function of $u$ only.
There exists therefore a (time-dependent) matrix
$C\=(C_i^a)\in GL(D-2,\IR)$ such that
$$
g_{ij}(u)=
\delta_{ab}C_i^a(u)C_j^b(u).
\equation
$$
Such a $C$ is unique up to an orthogonal matrix.
Then, introducing the inverse matrix $D\=C^{-1}$,
the coordinate transformation
$$
{\bf X}=C{\bf x},
\qquad
U=u,\qquad
V=v 
\equation
$$
flattens out the transverse metric
while preserving the form of
the other terms, i.e. yields
$$
d{\bf X}^2-2dU\big[dV
+\2E_{ij}X^jdX^i\big]
+K(U,{\bf X})dU^2,
\equation
$$
where
$$
\eqalign{
E&=(E_{ij})=
-2D^{-1}\dot{D}
+D^T\,e\,D,
\qquad
e=(e_{ij}),
\ccr
K&=k+{\bf X}^T\left[\dot{D}^T\big(D^{-1}\big)^T
D^{-1}\dot{D}
-\dot{D}^T\,e\,D\right]{\bf X},
\cr
}
\equation
$$
the superscript `$T$' denoting transposition. This generalizes
a result of Gibbons~[11].

Let us now decompose the matrix $E_{ij}(u)$ into the sum of a
symmetric and an antisymmetric matrix, $E_{ij}\=S_{ij}+A_{ij}$.

\kikezd{Step 2}. The symmetric part yields a gradient, 
$S_{ij}x^j\=\partial_i\Lambda$ with $\Lambda(u,{\bf x})
=\2S_{ij}(u)x^ix^j$, and can therefore be gauged away as 
$$
{\bf X}={\bf x}
\qquad
U=u
\qquad
V=v+\2\Lambda(u,{\bf x}),
\equation
$$
brings the metric (3.3) into the form
$$
d{\bf X}^2-2dU\big[dV
+\2A_{ij}X^jdX^i\big]
+K(U,{\bf X})dU^2,
\equation
$$
with
$$
K=k(U,{\bf X})+\partial_u\Lambda
=
k+\2\dot{S}_{ij}X^iX^j.
\equation
$$

\kikezd{Step 3}. The freedom in choosing the matrix $C$ can be 
used to transform away the antisymmetric part $A_{ij}$:
there exists an {\it orthogonal} time-dependent matrix
$\big(O_{ij}\big)\in O(D\-2)$ such that
$$
A\equiv(A_{ij})=-2\,O^{-1}\dot{O},
\equation
$$
and the coordinate transformation 
$$
{\bf X}=O{\bf x}
\qquad
U=u
\qquad
V=v
\equation
$$
gives the metric (3.6) the simple form (1.1) with
$$
K=k+\smallover1/4A_{i\ }^{\ k}A_{kj}X^iX^j.
\equation
$$

The absence of the conformal anomaly follows therefore from those
results proved in Refs~[3-5] for ordinary plane waves, provided
Eq.~(1.2) is satisfied. 

At this stage, we can include axions and dilatons: first, the metric
is brought into the ordinary plane wave form (1.1), and then the
condition for the vanishing of the anomaly is simply derived from
the Horowitz-Steif condition (1.3).

Observe that, at each step, the additional term added to the
coefficient of $k$ is quadratic in the transverse variable.
Thus, if the function $k$ in Eq.~(2.3) we started with was
quadratic,  we would end up with an exact plane wave and the
vanishing of the anomaly would follow non-perturbatively from
Refs.~[3] and~[5].

\goodbreak

\chapter{Examples.}

i) Consider first the special Brinkmann metric 
$$
\phi(u)\,d{\bf x}^2
-2du\Big[
dv+\2\,A_{ij}(u)x^jdx^i
-
\smallover1/4\dot\phi(u)\,\delta_{ij}x^jdx^i\Big]
+k(u,{\bf x})du^2,
\equation
$$
where
$A_{ij}$ is antisymmetric and $\phi>0$~[12,9]. As explained
in Section 2, this yields
an exact string vacuum as soon as the vacuum
Einstein equation ${\tilde R}_{uu}\=0$, i.e. 
$$
\Delta k-{A_{ij}A^{ij}\over2\phi} 
+{D\-2\over 2}\,(\log \phi)\ddot{\ }\phi=0,
\equation
$$
is satisfied.
Another way of obtaining this result is to carry out the
coordinate transformations indicated in Section 3,
$$
\left\{\matrix{
1.&{\bf X}=\sqrt{\phi}\,{\bf x},\hfill&U=u,&V=v\hfill
\ccr
2.&{\bf X}={\bf x},\hfill&U=u,
&V=v+\smallover1/8{\displaystyle\dot\phi\over\displaystyle\phi\,}
\,{\bf x}^2\hfill
\ccr
3.&{\bf X}=O\,{\bf x},\hfill
&U=u,&V=v\hfill.
\cr
}\right.
\equation
$$
where $-2O^T\,\dot{O}=(A_{ij})$. This results in expressing
(4.1) as the ordinary plane wave (1.1) with
$$
K(U,{\bf X})=k\big(u,{{\bf X}\over\sqrt{\phi}}\big)
-{1\over4\phi^2}A_{ik}A_{j\ }^{\ k}X^iX^j,
\equation
$$
and then Eq.~(4.2) follows simply from (1.2), $\Delta\,K\=0$.

Interestingly, the Ansatz (4.1) is built from the same
ingredients as the metric + axion + dilaton system studied by
Horowitz and Steif~[4,5,10], and the condition (4.2) is, {\it up
to the sign of the quadratic term} and to some re-definition of
the fields, the same as the condition (1.3) of Horowitz and
Steif~[4]. Adding to the metric (4.1) an axion and a dilaton,
$B_{ij}$ and $\Phi(u)$, the Horowitz-Steif condition (1.3) is
generalized to
$$
\Delta\,k
+\Big[{D-2\over2}\,(\log\phi)\ddot{\ }\phi+2\ddot{\Phi}\Big]
+\Big[\smallover1/{18}B_{ij}B^{ij}-{A_{ij}A^{ij}\over2\phi}\Big]
=0.
\equation
$$

Let us point out that Eq.~(4.5) hints also to a possible
cancellation between the vector potential and the 
axion. Using the non-symmeric connection-approach~[13], we have
shown recently~[14] that this is indeed the case.
For $\phi\=1,\Phi\=0$ for example, we recover a recent
result presented in Ref.~14.

ii) Things work similarly it Rudd's toroidal case (1.4).
The metric is plainly of the form
(2.3) and provides therefore an exact string vacuum as soon as
Einstein's equation,
${\tilde R}_{uu}\=\sum_i{\ddot{R}_i/R_i}\=0$,
is satisfied.
The more general condition (1.5) can be derived by
`straightening out' the transverse metric following Step~1 and
then by gauge-transforming,
$$
\left\{\matrix{
1.&X^i=2\pi R_i(u)x^i,\hfill
&U=u,&V=V\hfill
\ccr
2.&{\bf X}={\bf x},\hfill
&U=u,&V=v+\2\,\sum_i{\displaystyle\dot{R}_i
\over\displaystyle R_i\,}\,(x^i)^2\hfill
\cr
}\right.
\equation
$$
which
takes (1.4) into the exact plane wave
$$
d{\bf x}^2-2dudv+
\left(\sum_i{\ddot{R}_i\over R_i}\,({x^i})^2\right)du^2.
\equation
$$
The associated Einstein equation $\Delta K\=0$ is plainly the same
as ${\tilde R}_{uu}\=0$ above. Adding a dilaton, $\Phi(u)$,
condition (1.5) follows from the Horowitz-Steif condition (1.3).
The vanishing of the anomaly follows from the results in Ref.~[4]
perturbatively, and from~[5] non-perturbatively.
(Note that one could also add an axion in the same way.)

iii) The two previous Ans\"atze, (1.4) and (4.1), can be unified
by considering rather
$$
\phi(u)\,R_i^2(u) (dx^i)^2 
-2du\Big[
dv+\2A_{ij}(u)x^jdx^i
-\smallover1/4\dot{\phi}(u)\,R_i^2(u)
\,\delta_{ij}x^jdx^i
\Big]
+k(u,{\bf x})du^2.
\equation
$$
Repeating the previous calculation we get that
if
$$
\Delta k-{A_{ij}A^{ij}\over2\phi}
+\sum_i{\ddot{R}_i\over R_i}\,\phi
+{D-2\over2}\big(\log\phi\big)\ddot{\ }\phi=0,
\equation
$$
then the anomaly vanishes at all orders in
sigma-model perturbation theory.

\chapter{Discussion.}

In the `one-coupling case' considered in the first example of
Section~4, i.e. for $g_{ij}\=\phi(u)\,\delta_{ij}$, Step~1 can be
replaced by a conformal rescaling.
Indeed,
if ${\tilde g}_{\mu\nu}\=\phi(u)\,{\hat g}_{\mu\nu}$
then (see, e.g.~[15])
$$
{\tilde R}_{uu}
=
{\hat R}_{uu}
+{D-2\over2}\Big[
(\log\phi)\ddot{\ }+\2\left({\dot\phi\over\phi}\right)^2
\Big].
\equation
$$
Calculating the Ricci tensor of the rescaled metric 
and using (5.1) we, again, get the constraint (4.2).

Our Ansatz (2.3) is the most general pp-wave in
$D\=4$ dimensions~[7].
In higher dimensions this is no longer true, however, and (1.6) is
indeed more general than the ordinary plane wave (1.1).

In suitable coordinates all Brinkmann metrics~[7] can be written
as $g_{ij}(u,{\bf x})dx^idx^j-2dudv$, 
so that all information is encoded in the transverse metric.
In Section~3 we have `flattened out' the transverse metric under
the assumption that this latter is a function of $u$ only.
This is, however, {\it not} the most general case when higher-order
correction terms vanish.
Consider, for example,
the time {\it and} space dependent metric
$$
\sum_{i=1}^{D-2}{
\cosh^2\left(\sqrt{\epsilon}_i w_i(x^i+u)\right) (dx^i)^2
}
-2dudv,
\equation
$$
where $\epsilon_i\=\pm1$ and $w_i=\const$.
Introducing
$$
\left\{\eqalign{
&X^i={1\over\sqrt{\epsilon_i}w_i}
\sinh\left(\sqrt{\epsilon_i}w_i(x^i+u)\right),
\ccr
&V=v+\2\sum_i{\left(
{1\over2\sqrt{\epsilon_i}w_i}
\sinh\left(2\sqrt{\epsilon_i}w_i(x^i+u)\right)+x^i+u
\right)},
\ccr
&U=u,
\cr
}\right.
\equation
$$
then (5.2) is turned into the exact plane wave (1.1) with
$K\=\sum_i{\left(1+\epsilon_iw_i^2(X^i)^2\right)}$, which
is an exact string vacuum as soon as the vacuum Einstein
equation $\sum_i{\epsilon_iw_i^2}\=0$ is satisfied.

More generally, Step 1 can be implemented whenever the transverse
metric is conformally flat, which requires the transverse 
Weyl tensor to vanish for each fixed value of~$u$.

At last, it would be interesting to know 
(i) precisely when can a general Brinkmann metric (1.6) be brought 
into the ordinary plane-wave form (1.1)
and
(ii) if this is indeed necessary for the vanishing of all
higher-order terms.

\kikezd{Note added}. After this paper was published, anomaly cancellation has been proved non-perturbatively
in a Wess-Zumino context [16].

\kikezd{Acknowledgements}.
Parts of the results presented here are contained in an unpublished
note [12]. We are indebted to Gary Gibbons and Malcolm Perry for
their interest and collaboration at the early stages of this work.
After our paper was completed, we discovered that similar results
were found, independently, by Arkady Tseytlin, to whom we are
indebted for correspondence.
One of us (Z.~H.) would like to thank Tours University for 
hospitality extended to him, and to the Hungarian National Science
and Research Foundation (Grant No. 2177) for a partial financial
support.

\vskip 0.3cm

\goodbreak

%%%%%%%%%%%%%%%%%%%%%%%%%%%%%%%%%%%%%%%%%%%
%%%%%%%%%%%%% the references %%%%%%%%%%%%%%
%%%%%%%%%%%%%%%%%%%%%%%%%%%%%%%%%%%%%%%%%%%

\centerline{\bf\BBig References}

\reference
C.~Lovelace, Phys. Lett. {\bf B138}, 75 (1984);
E.~Fradkin and A.~A.~Tseytlin, Phys. Lett. {\bf B158}, 316 (1985);
{\bf B160}, 69 (1985);
Nucl. Phys. {\bf B261}, 1 (1985);
C.~Callan, D.~Friedan, E.~Martinec and M.~Perry, 
Nucl. Phys. {\bf 262B}, 593 (1985);
H.~J.~de Vega, in
%{\it Strings in curved spacetimes}
%Lectures delivered at the
Erice School "String quantum gravity and physics at the 
Planck energy scale",
June 1992, Proc. ed. N.~Sanchez, World Scientific.

\reference
M.~B.~Green, J.~H.~Schwarz and E.~Witten, {\it Superstring Theory}, Vol. 1,
Cambridge University Press, (1987).

\reference
D.~Amati and C.~Klim\v{c}\'\i{k}, Phys. Lett. {\bf B219}, 443 (1989);
R.~G\"uven, Phys. Lett. {\bf B191}, 275 (1987);
H.~J.~de Vega and N.~Sanchez, Nucl. Phys. {\bf B317}, 706 (1989);
M.~E.~V.~Costa and H.~J.~de Vega,
Ann. Phys. {\bf 211}, 223 (1991).

\reference
G.~T.~Horowitz and A.~R.~Steif, Phys. Rev. Lett. {\bf 64} (1990), 260; 
Phys. Rev. {\bf D42}, 1950 (1990).

\reference
A.~R.~Steif, Phys. Rev. {\bf D42}, 2150 (1990).

\reference
R.~Rudd, Nucl. Phys. {\bf B352}, 489 (1991).

\reference
H.~W.~Brinkmann, 
Math. Ann. {\bf 94}, 119 (1925).

\reference
C.~Duval, G.~W.~Gibbons and P.~Horv\'athy,
Phys. Rev. {\bf D43}, 3907 (1991) [hep-th/0512188].

\reference
A.~A.~Tseytlin, Phys. Lett. {\bf B288}, 279 (1992);
Nucl. Phys. {\bf B390}, 153 (1993);
Phys. Rev. {\bf D47}, 3421 (1993).

\reference
G.~T.~Horowitz, in Proc. VIth Int. Superstring Workshop
{\it Strings'90}, Texas '90, Singapore, World Scientific (1991).

\reference
G.~W.~Gibbons,
Commun. Math. Phys. {\bf 45}, 191 (1975).

\reference
C.~Duval, G.~W.~Gibbons, P.~A.~Horv\'athy, and M.~J.~Perry,
unpublished (1991).

\reference
E.~Braaten, Phys. Rev. Lett. {\bf 53}, 1799 (1984); 
E.~Braaten, T.~L.~Curtright and C.~K.~Zachos, 
Nucl. Phys. {\bf B260}, 630 (1985);
H.~Osborn, Ann. Phys. {\bf 200}, 1 (1990).

\reference
C.~Duval, Z.~Horv\'ath and P.~A.~Horv\'athy, 
Phys. Lett. {\bf 313B}, 10 (1993)
[hep-th/0306059].

\reference
D.~Kramer, H.~Stephani, E.~Herlt, M.~McCallum,
{\it Exact solutions of Einstein's field equations},
Cambridge Univ. Press (1980).

\reference
C. Nappi and E. Witten,
Phys. Rev. Lett. {\bf 71}, 3751 (1993) 
[hep-th/9310112]
\vfill\eject

\bye